\documentclass{kapproc} % Computer Modern font calls

\usepackage{procps} 
\usepackage{procps} 

\usepackage[dvips]{graphicx}
\upperandlowercase

\kluwerbib % will produce this kind of bibliography entry:

\begin{document}

\articletitle{A new scenario for the origin of galactic warps}

%\articlesubtitle{This is an Article Subtitle}

\author{Yves Revaz}
\affil{Geneva Observatory}
\email{yves.revaz@obs.unige.ch}

\author{Daniel Pfenniger}
\affil{Geneva Observatory, University of Geneva}
\email{daniel.pfenniger@obs.unige.ch}

%%%%%%%%%%%%%%%%%%%%%%%%%%%%%%%%%%%%%%%%%%%%%%%%%%%%%%%%%%%%%%%%%%
\begin{abstract}
%%%%%%%%%%%%%%%%%%%%%%%%%%%%%%%%%%%%%%%%%%%%%%%%%%%%%%%%%%%%%%%%%%
Galactic warps represent an old unresolved problem, since the discovery,
at the end of the fifties, of the HI warp of the Milky Way.
In this paper, we propose a new scenario explaining a large fraction of the observed
optical warps. Based on N-body simulations, we show that realistic galactic disks, 
where the dark matter is essentially distributed in a disk, are subject 
to bending instabilities. S, U-shaped, as well as asymmetric warps are 
spontaneously generated and in some cases are long-lived.  
While this scenario presents the advantage of explaining the three observed
types of warps, it also brings new constraints on the dark matter distribution
in spiral galaxies. Finally, it gives us a unified picture of galaxies where galactic 
asymmetries, like bars, spirals and warps result from gravitational instabilities.
\end{abstract}

\begin{keywords}
galactic dynamics, warps, dark matter
\end{keywords}

%%%%%%%%%%%%%%%%%%%%%%%%%%%%%%%%%%%%%%%%%%%%%%%%%%%%%%%%%%%%%%%%%%
\section{Why do disk galaxies have often spirals and warps?}
%%%%%%%%%%%%%%%%%%%%%%%%%%%%%%%%%%%%%%%%%%%%%%%%%%%%%%%%%%%%%%%%%%
Here we propose that the fundamental reason why spiral galaxies present frequently warps is 
identical to the one why spirals have frequently spiral arms and bars: the dissipational 
component, gas-dust, brings these systems toward a critical state with respect to gravitational 
instabilities, producing spiral arms and bars in the radial direction, and
warps in the vertical direction. 
As long as gas dissipates the dynamical reheating induced by the instabilities 
compensates the cooling due to radiative losses in the gas.  The critical state is 
maintained and disk galaxies remain very responsive to perturbations.  

In such conditions any perturbations, be it
gas infall or interactions, trigger large  and long-lived responses of the disks 
in the form of grand design spirals or large warps.
Thus while external perturbations help galaxies to produce spirals and warps 
the fundamental cause why the disk response is large and slow lies in the marginally 
stable state of such systems reached by the secular energy losses. 

Consistent with this scenario is that gas poor disk galaxies, such as S0's possess 
typically only small, tightly wound spiral arms, and warps are rare.  When too little gas is 
left it is unable to regenerate gravitational instabilities. 

The interesting by-product of such a scenario is that the matter content
of galaxy disks is then constraint.   
Several  observational results, like spiral structures in HI disks (\cite{masset03}), 
dark matter and HI correlation (~\cite{bosma78,hoekstra01,pfenniger05}), 
star formation in low gas density regions 
(\cite{cuillandre01}) and more recently, diffuse gamma-rays in the Galaxy
(\cite{grenier05}) suggest that dark baryons are in galactic disks 
associated to HI and classical molecular clouds.
A substantial amount of matter in the disks would therefore
explain many aspects of disks galaxies, and in passing warps. 

Bending instabilities are Jeans like  gravitational instabilities 
that occur in flat systems and cause them to warp. 
\cite{toomre66} and \cite{araki85} have shown that 
an infinite slab of finite thickness may be unstable when the ratio of 
the vertical velocity dispersion $\sigma_z$ to the velocity dispersion in the plane $\sigma_{u}$ 
is less than $0.293$ (This value is called the Araki limit). 
In this work we use N-body simulations to quantify the effects produced 
by bending instabilities in self-gravitating disks of different thicknesses, 
and to quantifiy the amount of matter that bending instabilities require in
the disk in order to be unstable, and the amount of matter that is left in
the halo up to the disk radius.    

%%%%%%%%%%%%%%%%%%%%%%%%%%%%%%%%%%%%%%%%%%%%%%%%%%%%%%%%%%%%%%%%%%
\section{The heavy disk model}
%%%%%%%%%%%%%%%%%%%%%%%%%%%%%%%%%%%%%%%%%%%%%%%%%%%%%%%%%%%%%%%%%%
We have studied the spontaneous formation of warp resulting from bending instabilities in 9 different 
galactic models. All mass model are composed of a bulge, an exponential stellar
disk ($H_R=2.5$, $H_z=0.25\,\rm{kpc}$) and a heavy disk made of HI gas and dark matter 
proportional to it. 
With a respective mass fraction for the 3 components of $0.068$, $0.206$, $0.726$  the rotation curve 
is approximately flat up to $R=35\,\textrm{kpc}$.
The 9 models differ only by the thickness $h_{z0}$ and flaring $R_{f} $ of the dark matter component. 

The initial vertical velocity dispersion $\sigma_z$ is found by
satisfying the equilibrium solution of the stellar hydrodynamic
equation in cylindrical coordinates, separately for each components.
To set the azimuthal and radial dispersions, we have used the epicycle approximation
and the ratio between the radial and vertical epicyclic frequency $\kappa$ and $\nu$.
See \cite{revaz04} for more details on the models.

\begin{figure}[ht]
  \rotatebox{-90}{\includegraphics[width=4cm]{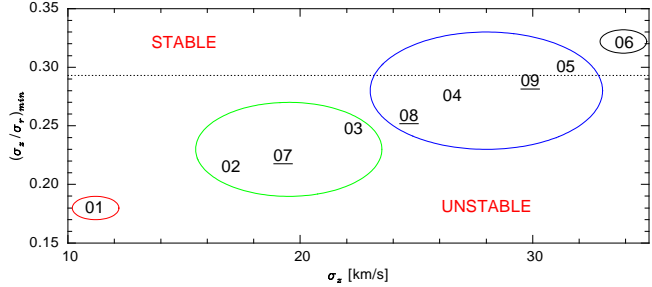}}
  \caption{Ratio $\sigma_z/\sigma_R$ as a function of the vertical
       dispersion $\sigma_z$ at $R=15\,\rm{kpc}$.  The values are taken at
       the radius where $\sigma_z/\sigma_R$ is minimum.  The dotted
       line corresponds to Araki's limit.}
  \label{szszr}
  \end{figure}

All models (from 01 to 09) are displayed on Fig.~\ref{szszr}, where the ratio $\sigma_z/\sigma_R$ 
(stability index) is plotted as a function of the vertical dispersion. 
This latter value gives an idea of the disk thickness.
Thin disks are found at lower left end while thick disks are found at the upper right end.  
The dotted line corresponds to Araki's stability criterion $\sigma_z/\sigma_R = 0.293$. We thus expect
that thinner disk will be less stable.
%

%%%%%%%%%%%%%%%%%%%%%%%%%%%%%%%%%%%%%%%%%%%%%%%%%%%%%%%%%%%%%%%%%%
\section{Stability as a function of the disk thickness}
%%%%%%%%%%%%%%%%%%%%%%%%%%%%%%%%%%%%%%%%%%%%%%%%%%%%%%%%%%%%%%%%%%

As expected from Fig.~2., the evolution of the models depends strongly on the
thickness of the heavy disk. According to their evolution, the models can be divided 
in four groups:
\begin{itemize}
\item[1)] Model 01 has a ratio $\sigma_z/\sigma_R$ of $0.18$.  It is very
unstable. The bending instability occurs quickly and generates a
transient asymmetric warp that extends up to $z=4\,\rm{kpc}$ at
$R=35\rm{kpc}$ (see Fig.~\ref{snapshots}(a)). After the perturbation the disk is thickened 
and the galaxy is stabilized.
\item[2)] Models 02, 07 and 03 have still a ratio $\sigma_z/\sigma_R$ well
below the Araki limit. The bending instability occurs during the first
$2\,$Gyr. An axisymmetric bowl mode $m=0$ (U-shaped warp) grows during about
$1\,\rm{Gyr}$ (Fig.~\ref{snapshots}(b)), before that $\sigma_z$ increases and stabilizes the
disk.
\item[3)] The four models 08, 04, 09 and 05 develop S-shaped warped modes
($m=1$).  Except model 05 which has a ratio
$\sigma_z/\sigma_R=0.3$ just above Araki's limit, all are unstable
with respect to a bending.  In the case of model 08, the warp is
long-lived and lasts more than $5.5\,\rm{Gyr}$, corresponding to about
5 rotation times at $R=30\,\rm{kpc}$ (Fig.~\ref{snapshots}(c)). 
\item[4)] Model 06 has a ratio $\sigma_z/\sigma_R$ well above $0.3$. 
The model is stable.
  \begin{figure}[ht]
  \rotatebox{-90}{\includegraphics[width=3cm]{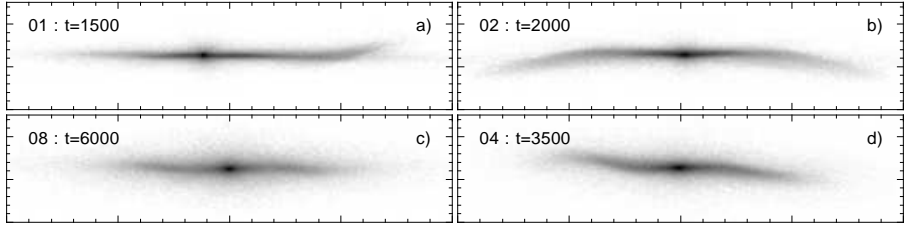}}
  \caption{Edge-on projections of the models 01, 02, 08 and 04.}
  \label{snapshots}
  \end{figure}

\end{itemize}
These simulations show that the bending instability in heavy disk model may be at the
origin of the 3 observed type of warp : S-shaped, U-shaped and asymmetric.
The high frequency of observed S-shaped warps (\cite{reshetnikov98}) is naturally explained 
because S-shaped warps develop near the Araki limit (marginally unstable regime) 
and then are long lived.

%%%%%%%%%%%%%%%%%%%%%%%%%%%%%%%%%%%%%%%%%%%%%%%%%%%%%%%%%%%%%%%%%%
\section{Dark matter constraint}
%%%%%%%%%%%%%%%%%%%%%%%%%%%%%%%%%%%%%%%%%%%%%%%%%%%%%%%%%%%%%%%%%%

The previous results have been obtained in the absence of a dark halo.
We have checked up to which point the disk self-gravity leads to bending 
instabilities when the disk is embedded in a conventional hot halo.

For a realistic model, where a fraction $(1-f)$ of the dark matter has been transfered 
into a halo, the vertical instabilities are damped if  
more than half of the dark mater is in the halo ($f<0.5$). On the contrary, the disk
may be unstable. 

When taking into account the dissipational behavior of the gas, as this is the case
in the plane for spirals, the disk is expected to reach a marginally unstable regime at
the Araki limit corresponding to $f=0.5$ in our model. 
We can then deduce, that half of the dark matter is situated in the disk 
and the other half in the halo. The flattening of our model is then in agreement with the flattening of 
the Milky Way, recently determined by \cite{johnston05} based on the study of the Sagittarius orbit.


\begin{chapthebibliography}{1}

\bibitem[Araki (1985)]{araki85}  
Araki, S.,
1985, Ph.D. Thesis, Massachusetts Inst. Technology

\bibitem[Bosma, 1978]{bosma78}                  
Bosma, A. 
1978, PhD thesis, Univ. Groningen

\bibitem[Cuillandre et al., 2001]{cuillandre01}         
Cuillandre, J-C., Lequeux, J., Allen, R.J., Mellier, Y., Bertin, E. 
2001, ApJ, 554, 190

\bibitem[Grenier et al., 2001]{grenier05}   
Grenier, I.A., Casandjian, J-M, Terrier, R.
2005, Science, 307, 1292
	
	
\bibitem[Hoekstra et al., 2001]{hoekstra01}     
Hoekstra, H., van Albada, T.S., Sancisi, R. 
2001, MNRAS, 323, 453  
	

\bibitem[Johnston et al. (2005)]{johnston05} 
Johnston, K.V., Law, D.R., Majewski, S.R.
2005, ApJ, 619, 800  

\bibitem[Masset \& Bureau, 2003]{masset03}      
Masset, F.D., Bureau, M.
2003, ApJ, 586, 152

\bibitem[Pfenniger \& Revaz, 2005]{pfenniger05}   
Pfenniger, D., Revaz, Y.
2005, A\&A, 431, 511

\bibitem[Reshetnikov, 1998]{reshetnikov98}      
Reshetnikov, V., Combes, F.
1998, A\&A, 337, 9

\bibitem[Revaz \& Pfenniger (2005)]{revaz04} 
Revaz, Y., Pfenniger, D. 
2004, A\&A,  425, 67

\bibitem[Toomre (1966)]{toomre66}        
Toomre, A.
1966, Geophys. Fluid Dyn., 46, 111



\end{chapthebibliography}
\end{document}